\documentclass[aps,prl,twocolumn,showpacs,superscriptaddress,groupedaddress, nofootinbib]{revtex4-1}

\usepackage{hyperref}
\usepackage{multirow}
\usepackage{mathrsfs,amsmath} 
\usepackage{verbatim}
\usepackage{amsmath}
\usepackage{amsfonts}
\usepackage{amssymb}
\usepackage{amsthm}
\usepackage{bm}
\usepackage{xparse}
\usepackage{graphicx}
\usepackage{xcolor}
\usepackage{textcase}
\usepackage{url}
\usepackage{lipsum}
\usepackage{appendix}
\usepackage{dsfont}
\usepackage{epstopdf}
\usepackage{footnote}
\usepackage{tgbonum}
\usepackage{blindtext}
\usepackage{enumitem}
\usepackage{mathrsfs,amsmath} 
\usepackage{float}
\usepackage{lipsum}
\usepackage{subfig}
\usepackage{subfloat}
\usepackage{wrapfig}
\usepackage{graphicx}
\usepackage{pslatex}
\usepackage{amsmath}
\usepackage{amsfonts}
\usepackage{setspace}
\usepackage{epstopdf}
\usepackage{wrapfig}
\usepackage{csquotes}
\usepackage{hyperref}
\usepackage{graphicx}
\usepackage{amssymb}
\usepackage{multirow}
\usepackage{xcolor}
\usepackage{framed}
\usepackage{mathtools}

\definecolor{shadecolor}{rgb}{0.75,0.75,0.75}
\DeclareDocumentCommand{\Tr}{m m O{\big}}{{\rm Tr}_{\:\!{#1}}#3({#2}#3)}

\newcommand{\R}{\mathbb{R}}
\newcommand{\Q}{\mathbb{Q}}
\newcommand{\N}{\mathbb{N}}

\begin{document}
\title{Physics without Determinism: Alternative Interpretations of Classical Physics}
\author{Flavio Del Santo}
\affiliation{
Institute for Quantum Optics and Quantum Information (IQOQI),
Austrian Academy of Sciences, Boltzmanngasse 3,
A-1090 Vienna, Austria and\\
Faculty of Physics,
University of Vienna,
Boltzmanngasse 5,
A-1090 Vienna, Austria}
\author{Nicolas Gisin}
\affiliation{Group of Applied Physics, University of Geneva, 1211 Geneva 4, Switzerland
}

\date{\today}

\begin{abstract}
Classical physics is generally regarded as deterministic, as opposed to quantum mechanics that is considered the first theory to have introduced genuine indeterminism into physics. We challenge this view by arguing that the alleged determinism of classical physics relies on the tacit, metaphysical assumption that there exists an actual value of every physical quantity, with its infinite predetermined digits (which we name \emph{principle of infinite precision}). Building on recent information-theoretic arguments showing that the principle of infinite precision (which translates into the attribution of a physical meaning to mathematical real numbers) leads to unphysical consequences, we consider possible alternative indeterministic interpretations of classical physics. We also link those to well-known interpretations of quantum mechanics. In particular, we propose a model of classical indeterminism based on \emph{finite information quantities} (FIQs). Moreover, we discuss the perspectives that an indeterministic physics could open (such as strong emergence), as well as some potential problematic issues. Finally, we make evident that any indeterministic interpretation of physics would have to deal with the problem of explaining how the indeterminate values become determinate, a problem known in the context of quantum mechanics as (part of) the ``quantum measurement problem''. We discuss some similarities between the classical and the quantum measurement problems, and propose ideas for possible solutions (e.g., ``collapse models'' and ``top-down causation'').

\end{abstract}

\maketitle
%%%%%%%%%%%%%%%%%%%%%%%%%%
\section{Why is classical physics (in)deterministic?}
It is generally accepted that classical physics (i.e., Newton's mechanics and Maxwell's electrodynamics) is \emph{deterministic}. Restating a famous argument due to Laplace (known as Laplace's demon), determinism is usually assumed to be the ``view that a sufficient knowledge of the laws of nature and appropriate boundary conditions will enable a superior intelligence to predict the future states of the physical world and to retrodict its past states with infinite precision'' \cite{determinism}.

Yet, stimulated by the development of statistical physics (which is taken to introduce indeterminacy as merely epistemic), one could find notable exceptions to this deterministic view, remarkably in the works of preeminent physicists the likes of L. Boltzmann, F. Exner, E. Schr{\"o}dinger and M. Born, who --admittedly with different standpoints-- all argued for genuine indeterminism in classical physics (see \cite{delsanto} and references thereof). These doubts about classical determinism were fostered in the second half of the twentieth century, after the theory of chaotic systems was systematically developed and its implications fully understood \cite{ornstein, prigogine}.

However, it is only in recent years that new life has been breathed into the critique of determinism in classical physics, showing that the advocacy of determinism leads to severe conceptual difficulties based on information-theoretic arguments \cite{dowek, gisin1, blundell}, and that determinism might even be incompatible with the derivation of the second law of thermodynamics \cite{drossel}. In fact, the hypothetical ``superior intelligence'' (demon), supposedly able to perfectly predict the future, is required to have complete information about the state of the universe and then use it to compute the subsequent evolution. Recent developments of information theory and its application to physics (mainly blossomed within the frameworks of quantum information and quantum thermodynamics) led to the conclusion that the abstract, mathematically well-formalized concept of information acquires a meaningful value in the natural sciences only if the information is embodied into a physical system (encoding), allowing it to be manipulated (computation) and transmitted (communication). As such, these processes are subject to the same limitations imposed by the laws of physics (\emph{Landauer's principle} \cite{landauer}). In the face of this, Laplace's demon, and hence determinism, leads to two categories of problems: the problem of \emph{infinities} and the problem of \emph{infinitesimals}.\footnote{Similar problems have  been recently discussed in a more general context and without resort to information-theoretic arguments in Ref. \cite{ellisinf}.}

The former of these problems is directly related to the memory capability of the physical systems that are supposed to encode the information of the whole Universe, then to be manipulated to compute the subsequent evolution. About this, Blundell concludes: ``If such a demon were (even hypothetically) to be constructed in our physical world, it would be subject to physical constraints which would include a limit on the number of atoms it could contain, bounded from above by the number of particles in the observable Universe. [...] Hence there is insufficient physical resource in the entire Universe to allow for the operation of a Laplacian demon able to analyze even a relatively limited macroscopic physical system'' \cite{blundell}.\footnote{The problem with (physical) infinities is connected to the so-called \emph{Hilbert's Hotel paradox}, proposed by D. Hilbert in 1924 \cite{hilbert}. This paradox illustrates the possibility for a hypothetical hotel with (countably) infinite rooms, all of which already occupied, to allocate (countably) infinitely many more guests.}

The problem of infinitesimals, instead, is related to the question of whether it would be possible to know, even in principle, the necessary boundary conditions with infinite precision.\footnote{In what follows, boundary and initial conditions will be used interchangeably, for they are conceptually the same in our discussion. In fact, they both serve as necessary inputs to the (differential) dynamical equations in order to give predictions. Thus, if either of the initial or the boundary conditions are not determined with infinite precision, they affect the subsequent dynamics in the same way.} Moreover, do these infinite-precision conditions (i.e., with infinite predetermined digits) exist at all?  As Drossel pointed out, this is related to the problem of determinism in so far as the  ``idea of a deterministic time evolution represented by a trajectory in phase space can only be upheld within the framework of classical mechanics if a point in phase space has infinite precision'' \cite{drossel}. To address the problem of infinitesimals, and in doing so challenging determinism both at the epistemic and ontic levels, one can again use an argument that relates information and physics, namely the fact that finite volumes can contain only a finite amount of information (\emph{Bekenstein bound}, see \cite{gisin1}).

In this respect, one should realize that classical physics is not inherently deterministic just because its formalism is a set of deterministic functions (differential equations), but rather its alleged deterministic character is based on the metaphysical, unwarranted assumption of ``infinite precision''. Such a hidden assumption can be formulated as a principle  --tacitly assumed in classical physics-- which consists of two different aspects:

\textbf{\emph{Principle of infinite precision}}:\\
1. (Ontological) -- there exists an actual value of every physical quantity, with its infinite determined digits (in any arbitrary numerical base).\\
2. (Epistemological) -- despite it might not be possible to know all the digits of a physical quantity (through measurements), it is possible to know an arbitrarily large number of digits. 

In this paper, we further develop the argument put forward in Ref. \cite{gisin1}, wherein it has been outlined a concrete possibility to replace the \emph{principle of infinite precision}. According to this view, the limits of this principle rely on the faulty assumption of granting a physical significance to mathematical real numbers. We would like to stress that such an assumption cannot be whatsoever justified at the operational level, as already stressed by Born, as early as 1955: ``Statements like `a quantity x has a completely definite value' (expressed by a real number and represented by a point in the mathematical continuum) seem to me to have no physical meaning" \cite{born}. Relaxing the assumption of the physical significance of mathematical real numbers, allows one to regard classical physics as a fundamentally indeterministic theory, contrarily to its standard formulation. The latter can be considered, in this view, a deterministic completion in terms of (tacitly) posited \emph{hidden variables}. This situation resembles (without however being completely analogous) the contraposition between the standard formulation of quantum mechanics --which considers indeterminism an irrefutable part of quantum theory-- and Bohm's \cite{bohm} or Gudder's \cite{gudder} hidden variable models, which provide a deterministic description of quantum phenomena --adding in principle inaccessible supplementary variables.

Before further discussing, in what follows, the arguments for alternative interpretations of classical physics without real numbers --and the issues that this can cause-- some general remarks on indeterminism seem due. It appears, in fact, that a common misconception concerning physical indeterminism, is that --in the eyes of some physicists and philosophers-- this is taken to imply that any kind of regularity or predictive power looks unwarranted --the supporter of determinism would ask: how can you explain the incredible predictive success of our laws of physics without causal determinism? Yet, an indeterministic description of the world does not (at least necessarily) entail a ``lawless indeterminism'', namely a complete disorder devoid of \emph{any} laws or regularities (quantum mechanics with its probabilistic predictions provides a prime example of these indeterministic regularities). We would like to define indeterminism trough the sufficient condition of the existence of \emph{some} events that are not fully determined by their past states; in the words of K. Popper, ``indeterminism merely asserts that there exists at least one event (or perhaps, one kind of events [...]) which is not predetermined'' \cite{popper50}. Such a remark is important because, historically, classical mechanics allowed to predict with a tremendous precision the motion, for instance, of the planets in the Solar System and this led Laplace to formulate his ideas on determinism, which became the standard view among physicists and remained for a long time unchallenged. In fact, thinking that physics is deterministic seems completely legit, in so far as certain physical systems exhibit extremely stable dynamics --e.g. an harmonic oscillator (pendulum), or the (Newtonian) gravitational two-body problem, and in general any integrable systems.  However, this justification of determinism can be challenged on the basis of two considerations. On the one hand, as already remarked, the existence of \emph{some} very stable systems (for all practical purposes treated as deterministic) does not undermine the possibility of indeterminism in the natural world. On the other hand, in the last one century a systematic study of chaotic systems --which are not integrable-- has been carried out, giving us good reasons to doubt determinism. Indeed, chaotic systems are not stable under perturbations, meaning that an arbitrarily small change in the initial conditions would lead to a significantly different future behavior, thus making the principle of infinite precision even more operationally unjustified, and therefore representing a concrete challenge to determinism.\footnote{Sometimes a distinction is made between strong and weak determinism. The former can be intuitively defined as ``\emph{similar} initial conditions lead to \emph{similar} trajectories’ and it is fulfilled by any integrable system. On the other hand, weak determinism can be defined as ``\emph{identical} conditions lead to \emph{identical} trajectories’. This holds for classical chaotic systems, however it is not empirically testable, for it would require the knowledge of the initial conditions with infinite precision.}

Incidentally, it is interesting to stress that, in the context of quantum physics, Bell's inequalities \cite{bell} have given us good reasons to believe, if not directly in indeterminism --which indeed cannot be confirmed or disproved within the domain of science (see further)-- that having at least one not predetermined event in the history of the Universe could have tremendous consequences on the following evolution. In fact, an experimental violation of Bell’s inequalities guarantees that if the inputs (measurement settings) were independent of the physical state shared by two distant parties, then the outcomes would be genuinely random (i.e., they cannot have predetermined values). Yet, the amount of random numbers generated in a Bell's test can be greater than the number of the corresponding inputs (in terms of bits of information): Bell's tests performed on quantum entangled states can be thought of as “machines” to increase the amount of randomness in the Universe (see also \cite{gisin2010}). Surprisingly enough, recent results \cite{renner, putz} showed that it is not even necessary to have a single genuinely random bit from the outset, but it is sufficient to introduce an arbitrarily small amount of initial randomness (i.e., of measurement independence) to generate virtually unbounded randomness. Hence, if one single event in the past history of the Universe was not fully causally determined beforehand, there is an operationally well defined procedure that allows to arbitrarily multiply the amount of indeterministic events in the future. Namely, it would be enough to use the randomness of the one indeterminate event as input in a Bell's test to extract more randomness (through the violation of a Bell's inequality). The random outputs can be used as new inputs for more Bell's experiments, and the process can be repeated arbitrarily many times \cite{amplification}. However, the initial arbitrarily small amount of randomness (or of indeterministic events) cannot be demonstrated by physics and its justification can only come from metaphysical arguments (see \cite{suppes}).

%%%%%%%%%%%%%%%%%%%%%%%%%%%%%%%%%%%%%%%%%%%%%%%%%%
%%%%%%%%%%%%%%%%%%%%%%%%%%%%%%%%%%%%%%%%%%%%%%%%%%

\section{Forms of indeterminism in classical and quantum physics}
Before proposing some possible models of indeterministic classical physics, in this section we shortly discuss some general features of deterministic and indeterministic theories. In doing so we aim at clarifying possible similarities between classical and quantum physics. Both classical and quantum mechanics are, in fact, formalized by a set of differential equations (laws of motion) that govern the dynamics of systems, together with appropriate initial conditions (IC) that fix the free parameters of these equations. Thus, if one aims at eliminating determinism as an unfounded interpretational element, there seems to be different possibilities, either involving the laws of physics or the characterization of the IC:

\textbf{1.} \emph{The laws of motions are fundamentally stochastic}. In this case, however, we cannot speak of an interpretation of the theory, but an actual modification of the formalism is required. In fact, in this case not only chaotic systems but also integrable ones would exhibit noisy outcomes, leading to experimentally inequivalent predictions. This case is the analogue of spontaneous collapse theories in quantum mechanics \cite{grw, gisincollapse, belljumps, cls}, which modify the Schr{\"o}dinger's equation with additional non-unitary terms.

\textbf{2.} \emph{The IC cannot, in principle, be fully known}. Without any ontological commitments, one can take seriously the epistemological statements of the \emph{principle of infinite precision} and push it to the extreme, namely asserting that there are in principle limits to the possibility of knowing (or measuring) certain quantities. This is what is entailed by the standard interpretation of the Heisenberg uncertainty principle that is usually stated as: there is in quantum physics a fundamental limit to the precision with which canonically conjugated variables \emph{can be known}.  Such an uncertainty can be introduced in classical physics as well, and would be characterized by a new natural constant $\epsilon$ (e.g., the standard deviation of a Gaussian function centered in the considered point). This would set an epistemic (yet fundamental) limit of precision, with which physical variables could be determined, as a classical analogue of  Heisenberg's uncertainty relations in quantum theory. This viewpoint appears similar to the one proposed in Ref. \cite{drossel}. Notice, however, that since this approach is agnostic with respect to the underlying ontology, it is fully compatible with a realist position that takes this uncertainty as being an ontological indeterminacy.
\footnote{As a supporting argument for this fundamental epistemic limit, one can even think in terms of theoretical reduction (i.e. when a theory is supervened by a more fundamental one, of which it represent an approximation). In fact, determining positions in classical physics with higher and higher precision, means to access digits that are relevant at the microscopic scale, and thus the Heisenberg's indeterminacy relations need to be applied (leading to the identification $\epsilon=\hbar$). However, it ought to be remarked that in certain classical chaotic systems the digits that are to become relevant are not necessarily the ones in the microscopic domain, but could be those which will become relevant only after a longer time} (such as in weather forecasts). So more general arguments than theoretical reduction would be desirable.

\textbf{3.} \emph{The IC \emph{are} not fully determined}. One can think that the fundamental limit of precision in determining physical quantities is not merely epistemic, but actually is an objective, ontologic indeterminacy that depends on the system and its interactions at a certain time. This view is the one that will be pursued in what follows, when we will propose ways to remove real numbers $\R$ from the domain of physics. Although this case does not seem to be the analogue of any specific interpretation of quantum theory, it clearly goes in the direction of realistic interpretations. It ought to be stressed, however, that if the initial conditions are not fully determined, this even makes (quantum) Bohmian mechanics becoming indeterministic.

\section{Indeterministic classical physics without real numbers}
In Ref.  \cite{gisin1}, a model of indeterministic classical mechanics has been sketched, which, while leaving  the dynamical equations unchanged, proposes a critical revision of the assumptions on the initial condition. In this view, the standard interpretation of classical mechanics has always tacitly assumed as  hidden variables the predetermined values taken by physical quantities in the domain of mathematical real numbers,  $\R $.  The physical existence of an infinite amount of predetermined digits could lead to  unphysical situations, such as the aforementioned  infinite information density, as explained in detail in Refs. \cite{dowek} and \cite{gisin1}.

In this section, we discuss some possible solutions to eliminate these unwanted features, namely possible ways of carrying out the relaxation of the postulate according to which physical quantities take values in the real numbers. These solutions are however intended to be merely different interpretations of classical mechanics, i.e. they ought to be empirically indistinguishable from the standard predictions (in the same way as interpretations of quantum mechanics are) \cite{baumann}.

\subsection{1. ``Truncated real numbers''.} A first possibility is to consider physical variables as taking values in a set of ``truncated real numbers''. This, as already noted by Born, would ensure the empirical indistinguishability from the standard classical physics: ``a statement like $x = \pi$ cm would have a physical meaning only if one could distinguish between it and $x = \pi_n$ cm for every $n$, where $\pi_n$ is the approximation of $\pi$ by the first $n$ decimals. This, however, is impossible; and even if we suppose that the accuracy of measurement will be increased in the future, $n$ can always be chosen so large that no experimental distinction is possible" \cite{born}. If one, however, wishes to attribute an ontological value to such an interpretation has to identify  $n$ with a new universal constant, that, independent of how big it could be, sets a limitation to the length of physically significant numbers, ultimately including the life of the Universe, if time too is to be considered a physical quantity. Another problematic issue is that $n$ would be dependent on the units in which one expresses the considered physical variables. This leads to consider a second possible solution.

\subsection{2. Rational numbers.} Another possibility is to consider that physical quantities take value in the rational numbers, $\Q$. Even if this sounds somewhat strange, one can argue that, in practice, physical measurements are in fact only described by rational numbers. For instance, a measurement of length is obtained by comparing a rod that has been carefully divided into equal parts (i.e. a ruler) with the object to be measured and determining the best fit within its (rational) divisions. And even probabilities are obtained as limits of frequencies of events' occurrences  (i.e. ratios of counts). However, while rational numbers do eliminate the unwanted infinite information density, they do not seem to remove determinism in so far as \emph{all} the digits are fully predetermined. Moreover, the use of rational numbers leads to those that can be named ``Pitagora's no-go theorems". Indeed, positing a physics based on rational numbers, would rule out the possibility of constructing a physical object with the shape of a perfect square with unit edge or a perfect circle with unit diameter. In fact, by means of elementary mathematical theorems, their diagonal  and circumference, respectively, would measure $\sqrt2$ and $\pi$, hence resulting to be physically unacceptable.
Additionally, if one plugs in the equations of motion initial conditions and time both taking values in the rational numbers, the solutions are not in general rational numbers.
These problematic issues lead to consider yet another possible solution.

\subsection{3. ``Computable real numbers''.}
A further alternative is to substitute the domain of physically meaningful numbers from (mathematically) real numbers to the proper subset thereof of ``computable real numbers''; that is, to keep all real numbers deprived of  the irrational, uncomputable ones. In fact, even irrational, computable real numbers can encode at most the same amount of non-trivial information (in bits) as the length of the shortest algorithm used to output their bits (i.e., the \emph{Kolmogorov complexity}). Uncomputable real numbers are in this model instead substituted by genuinely random numbers, thus introducing fundamental randomness also in classical physics. These numbers, together with chaotic systems, lay the foundations of an alternative classical indeterministic physics, which removes the paradox of infinite information density. However, this proposal could be considered an \emph{ad hoc} solution, since it maintains a field of mathematical numbers as physically significant, but removes ``by hand'' those that are problematic (which admittedly are almost all).

\subsection{4. ``Finite information quantities'' (FIQs).} 
Developing further the proposal in \cite{gisin1}, we put forward an alternative class of random numbers which are for all practical purposes (in terms of empirical predictions) equivalent to real numbers, but that have actually zero overlap with them (they are not a mathematical number field, nor a proper subset thereof). We refer to them as ``finite-information quantities'' (FIQs). In order to illustrate this possible alternative solution to overcome the problems with the principle of infinite precision, let us consider again the standard interpretation in greater formal detail. A physical quantity $\gamma$ (which may be the scalar parameter time, a universal constant, as well as a one-dimensional component of the position or of the momentum, etc.) is assumed to take values in the domain of real numbers, i.e., $\gamma \in \R$. Without loss of generality, but as a matter of simplicity, let us consider $\gamma$ to be between 0 and 1, and that its digits (bits) are expressed in binary base:
\begin{equation*}
\gamma=0.\gamma_1\gamma_2\cdots \gamma_j \cdots,
\end{equation*}
where each $\gamma_j\in\{0,1\}$, $\forall j\in \N^+$. This means that, being $\gamma \in \R$, its infinite bits are \emph{all} given at once and each one of them takes as a value either 0 or 1. 

In an indeterministic world, however, not all the digits should be determined at all times, yet we require this model to give the same empirical predictions of the standard one. We therefore require a physical quantity to have the first (more significant) $N$ digits fully determined --and to be the same as those that give the standard deterministic predictions-- at time $t$, and we write $\gamma(N(t))$, whereas the following infinite digits are not yet determined. This reads:
\begin{equation*}
\gamma \left(N(t)\right)=0.\gamma_1\gamma_2\cdots \gamma_{N(t)} ?_{N(t)+1}\cdots ?_k\cdots,
\end{equation*}
where each $\gamma_j\in\{0,1\}$, $\forall j\leq N(t)$, and the symbol $?_k$ here means that the $k$th digit is a not yet actualized binary digit (see further).

Despite the element of randomness introduced, the transition between the actualized values and the random values still to be realized does not need to be a sharp one. In fact, one can conceive an objective property that quantifies the (possibly unbalanced) disposition of every digit to take one of the two possible values, 0 or 1. This property is reminiscent of Popper's propensities \cite{popper},\footnote{\label{note}While propensities were for Popper an interpretation of mathematical probabilities proper, we are not here necessarily requiring them to satisfy Kolmogorov's axioms, as discussed in Ref. \cite{gisin2}.} and it can be seen as the element of objective reality of this alternative interpretation:

\textbf{Definition - \textit{propensities}}\\
There exist (in the sense of  being ontologically real) physical properties that we call \emph{propensities} $q_j\in [0,1] \cap \Q$, for each digit $j$ of a physical quantity $\gamma(N(t))$. A propensity quantifies the tendency or disposition of the $j$th binary digit to take the value 1.

{The interpretation of propensities can be understood starting from the limit cases. If the propensities are 0 or 1 the meaning is straightforward. For example, $q_j=1$ means that the $j$th digit will take value 1 with certainty. On the opposite extreme, if a bit has an associated propensity of 1/2, it means that the bit is totally random. Namely, if one were to measure the value of this bit, there would be an intrinsic property that makes it taking the value 0 or 1 with equal likelihood (we don't use ``probability'' to avoid formal issues, see footnote \ref{note}). All the intermediate cases can then be constructed. For instance, a propensity $q_k=0.3$ means that there is an objective tendency of the $k$th digit to take the value 1, quantified by 0.3, and thus the complementary propensity of taking the value 0 would be 0.7 (how this actualization occurs is an open issue, as we discuss in the next section). We would like to stress that while we assume propensities to be an (ontic) objective property, at the operational level they lead to the measured (epistemic) frequencies, but they supervene frequencies insofar as propensities can describe single-time events.}

{red}{ We posit that propensities take values in the domain of rational numbers such that they contain only a finite amount of information. Hence, postulating them as an element of reality, does not lead to the same information paradoxes of real numbers.} It also follows from the definition, that the propensities $q_j$ for the first $N(t)$ digits of a quantity $\gamma(N(t))$ at time $t$ are all either 0 or 1, i.e. $q_j\in\{0,1\}$, $\forall j\in[1,N(t)]$.

As a function of time, propensities must undergo a dynamical evolution. We envision more than a way to evolve propensities in time, although we do not propose an explicit model to describe this. On the one hand, one can think of a dynamical process similar to spontaneous collapse models of quantum mechanics. Admittedly, spontaneous collapse models require to modify the fundamental dynamical equation of quantum physics, the Schr{\"o}dinger's equation. Hence these models are not merely interpretations, but testable different theories. Nevertheless, for propensity this is not necessarily the case because they are a postulated element of reality which is however not observable. Thus, even if it would be desirable to have an explicit form for the equations governing the dynamics of propensities, the measured values of physical (observable) quantities would evolve in the usual way. Thus we maintain that our proposed new interpretation is indeed an interpretation and not a different testable theory.

On the other hand, intuitionistic mathematics could be the tool to solve the issue of the evolution of propensities (see ``choice sequences'' below). In fact, one can start from an infinite sequence of completely random bits (or digits), and then the number representing a physical quantity evolves according to a law (a function of these random bits). However, despite this law would describe the evolution of propensities, it is different from a standard a physical law, for it is a different way to construct mathematical numbers --which in turn describe physical quantities-- in time.

Making use of propensities, we can now refine our definition of physical quantities:

\textbf{Definition - \textit{FIQs}}\\
A \emph{finite-information quantity} (FIQ) is an ordered list of propensities $\{ q_1, q_2, \cdots , q_j, \cdots \}$, that satisfies:\\
1. (necessary condition): The information content is finite, i.e.  $\sum_j I_j < \infty$, where $I_j=1-H(q_j)$ is the information content of the propensity, and $H$ is the binary entropy function of its argument. This ensures that the information content of FIQs is bounded from above;\\
2. (sufficient condition): After a certain threshold, all the bits are completely random, i.e. $\exists M(t) \in \N$ such that $q_j = \frac{1}{2}, \ \ \forall j>M(t)$\\

It ought to be stressed that this view grants a prior fundamentality to the potential property of becoming actual (a list of propensities, FIQ), more that to the already actualized number (a list of determined bits). {In fact, the analogue of a \emph{pure state} in this alternative interpretation of classical physics would be a collection of all the FIQs associated with the dynamical variable (i.e., the list of the propensities of each digit). Namely, this represents the maximal piece of information regarding a physical system. Yet, even having access to this knowledge (which is admittedly not possible due to the fact that propensities are not measurable) would lead to in principle unpredictable different evolutions. Thus, two systems that are identical at a certain instant of time (in the sense that they are in the same pure state, i.e. the propensities associated to their variables are all the same) will have, in general, different observable behaviors at later times.} However, the merit of this view is that the bits are realized univocally and irreversibly as time passes, but the information content of a FIQ is always bounded, contrarily to that of a real number. A physical quantity $\gamma$ reads in this interpretation as follows:
\begin{equation*}
\gamma \left(N(t), M(t)\right)=0.\underbrace{\gamma_1\gamma_2\cdots \gamma_{N(t)}}_{\textrm{determined }\gamma_j\in \{0, 1\}} \overbrace{?_{N(t)+1}\cdots ?_{M(t)}}^{?_k\textrm{, with } q_k\in(0, 1)}\underbrace{?_{M(t)+1}\cdots}_{?_l\textrm{, with } q_l=\frac{1}{2}}.
\end{equation*}
Notice that none of the FIQs is a mathematical number, but they capture the tendency (propensity) of each bit of a physical quantity to take the value 0 or 1 at the following instant in time. This admittedly leads to problematic issues, such as the problem of how and when the actualization of the digits from their propensity take place: it thus introduces the analogue of the quantum measurement problem also in classical physics (see further).

Moreover, FIQs partly even out the fundamental differences between classical and quantum physics, making both of them indeterministic (and making so even Bohmian interpretation).\footnote{There is yet another deterministic interpretation of quantum mechanics, the so-called \emph{many-worlds interpretation} that grants physical reality to the wave function of the Universe, which always evolve unitarily. If FIQs are introduced in that interpretation, then the realization of \emph{all} the values of the bits would actually take place, each of which being real in a different ``world''.} Table \ref{table} compares some possible combinations of deterministic and indeterministic interpretations of quantum and classical physics.
\begin{table*}[]
\centering

\begin{tabular}{c|c|c|}
\cline{2-3}
                                                       & Classical                                                                                   & Quantum                                                                                                                                                                              \\ \hline
\multicolumn{1}{|c|}{\multirow{2}{*}{\rotatebox[origin=c]{90}{Indeterministic}}} & \multirow{2}{*}{\begin{tabular}[c]{@{}c@{}}IC $\in$ FIQs\\ Newton's equation\end{tabular}} & \begin{tabular}[c]{@{}c@{}}$|\psi\rangle \in \mathcal{L}^2(\R^N)$ \\Measurement postulate\end{tabular}                         \\ \cline{3-3} 
\multicolumn{1}{|c|}{}                                 &                                                                                             & \begin{tabular}[c]{@{}c@{}}IC (position) $\in$ FIQs and $|\psi\rangle \in \mathcal{L}^2(\R^N)$\\ Bohm's guidance equation\\ admitted\end{tabular} \\ \hline
\multicolumn{1}{|c|}{\multirow{2}{*}{\rotatebox[origin=c]{90}{Deterministic}}}   & \multirow{2}{*}{\begin{tabular}[c]{@{}c@{}}IC $\in \R$\\ Newton's equation\end{tabular}}  & \begin{tabular}[c]{@{}c@{}}IC (position) $\in \R$ and $|\psi\rangle \in \mathcal{L}^2(\R^N)$\\ Bohm's guidance equation\\ Schr{\"o}dinger's equation\end{tabular}  \\ \cline{3-3} 
\multicolumn{1}{|c|}{}                                 &                                                                                             & Many worlds interpretation (?)                                                                                                                                                       \\ \hline
\end{tabular}

\caption{\small{A table comparing deterministic and indeterministic interpretations of classical and quantum physics. Note that the substitution of FIQs in the place of real numbers makes not only classical physics indeterministic, but also  Bohm's interpretation of quantum physics (which is usually taken to restore determinism).}}
\label{table}
\end{table*}

\subsection{5. ``Choice sequences''.}
Finite Information Quantities are not numbers in the usual sense, because their digits are not all given at once. On the contrary, the ``bits" of FIQs evolve as time passes, they start from the value $\frac{1}{2}$ and evolve until they acquire a bit value of either 0 or 1. In a nutshell, FIQs are processes that develop in time. Interestingly, in intuitionistic mathematics, the continuum is filled by ``choice sequences", as Brouwer, the father of intuitionism, and followers named them \cite{Brouwer1948}. This is not the place to present intuitionistic mathematics (see, e.g., \cite{IndeterminateNumbersPosy}), but let us emphasize that this alternative to classical (Platonistic) mathematics allows one to formalize ``dynamical numbers" that resemble much our FIQs \cite{Troelstra}. Interestingly, using the language of intuitionistic mathematics makes it much easier to talk of indeterminism \cite{NGHiddenReals}.

\section{Strong emergence}
The argument for determinism seems to rely, to a certain extent, on the tacit assumption of reductionism in its stronger form of microphysicalism, i.e. the view that every entity and phenomenon are ultimately reducible to fundamental interactions between elementary building blocs of physics (e.g., particles). In fact, in a completely deterministic picture, every particular phenomenon can be traced back to the interactions between its primitive components, along a (finite) chain of causally predetermined events. In this way any form of strong emergence seems to be ruled out, and it becomes only apparent (i.e. a weak or epistemic emergence). On the other hand, admitting genuine randomness in the universe, allows in our opinion the possibility of strong emergence.\footnote{For a definition of strong emergence, see, for instance, Ref. \cite{chalmers}: ``We can say that a high-level phenomenon is strongly emergent with respect to a
low-level domain when the high-level phenomenon arises from the low-level domain, but truths concerning that phenomenon are not deducible even in principle from truths in the low-level domain''.}\\
As a concrete example, consider the kinetic theory of gases. If one starts from a molecular description of the ideal gas, from the perspective of standard, deterministic classical mechanics, the stochasticity is only epistemic (i.e. only an apparent effect due to the lack of complete information regarding positions and momenta of every single molecule). Thus, the deterministic behavior of the law of the ideal gas is not expected to be a strong emergent feature, but solely a retrieving at the macroscopic scale of the fundamental determinism of the microscopic components.
In the perspective of the alternative indeterministic interpretation (based on FIQs), instead, the deterministic law of the ideal gas, ruling the behavior at the macroscopic level, emerges as a novel and not reducible feature, from fundamental randomness.\footnote{It is true that also the law of the ideal gas would not be \emph{perfectly} deterministic in a FIQ-based physics, however its stability makes it almost deterministic for all practical purposes, whereas at the microscopic level, chaotic behaviors multiply the fundamental uncertainty of the single molecules.}

Notice that the historical debate on the apparent incompatibility between Poincar\'e's recurrence theorem and Boltzmann's kinetic theory of gases does not arise in the framework of FIQs. Poincar\'e's recurrence theorem, in fact, states that continuous-state systems (i.e., in which the state variables change continuously in time) return to an arbitrarily small neighborhood of the initial state in phase space. However, Poincar\'e's theorem relies on the fact that the initial state is perfectly determined (i.e., it is a mathematical point identified by a set of coordinates which take values in the real numbers) in phase space. Thus in a FIQ-based alternative physics the theorem simply cannot be derived. In fact, FIQs interpretation features genuinely irreversible physical processes. 

Similarly, Drossel has recently pointed out that, in a physics where it is impossible to determine points of phase space with infinite precision, ``the time evolution of thermodynamics is undetermined by classical mechanics [...]. Thus, the second law of thermodynamics is an emergent law in the strong sense; it is not contained in the microscopic laws of classical mechanics'' \cite{drossel}.

%\textcolor{red}{Is there more than merely the law of large numbers? Even if we can’t answer this we should dare to raise the question and suggest answers.}

At this point one should ask oneself whether there are examples of emergence that possibly go beyond some form of the law of large numbers. Admittedly, we are unsure about this. Clearly, in all indeterministic physical theories, the law of large numbers will play an important role and lead to some stability and hence to some form of determinism at the larger scale (or higher-level description). It seems that this question is closely related to possible top-down causation, the topic of the next section.

\section{Top-down causation}
The idea of strong emergence, including emergent determinism, is related to the concept of ``top-down causation'' \cite{topdown, topdown2}. In this view, microphysicalism is not necessarily rejected ontologically (i.e., it admits that complex structures are hierarchical modular compositions of simpler ones), but the fact that the behavior of macroscopic events is fully determined by the interactions of the macroscopic entities is revised.  Top-down causation maintains that the interactions between microscopic entities do not causally supervene the macroscopic phenomena, but rather it posits a mutual interaction where also the macroscopic (strongly emergent) laws impose constraints on the behavior of their constituents. Note that top-down causation requires indeterminism (at least at the lower level of the constituents) to be in principle conceivable \cite{drosselnew}; this was already remarked by Popper, when stating: ``{[A] higher level may exert a dominant influence upon a lower level}. For it seems that, were the universe \emph{per impossibile} a perfect determinist clockwork, there would be no heat production and no layers and therefore no such dominating influence would occur. This suggests that the emergence of hierarchical levels or layers, and of an interaction between them, depends upon a fundamental indeterminism of the physical universe. Each level is open to causal influences coming from lower and from higher levels'' \cite{poppernew}.

Concerning indeterministic interpretations of physical theories, top-down causation could help to understand how the determination of dynamical variables (i.e., the actualization of their values) occurs in the context of indeterministic theories. Namely, the reason why --and under what circumstances-- a single definite value is realized among all the possible ones. In the FIQ-based indeterministic interpretation of classical physics here introduced, this translates into the understanding of how the bits of physical variables becomes fully determined, namely how their propensities become either 0 or 1. We envision two possible mechanisms that could explain the actualization of the variables:

\textit{1. The actualizations happens spontaneously as time passes}. This view is compatible with reductionism and it does not necessarily require any effects of top-down causation. Note that this mechanism resembles, in the context of quantum mechanics, objective collapse models such as the ``continuous spontaneous localization'' (CSL) \cite{gisincollapse, cls}. 

\textit{2. The actualization happens when a higher level requires it}. This means that when a higher level of description (e.g., the macroscopic measurement apparatus) requires some physical quantity pertaining to the lower-level description to acquire a determined value, then the lower level must get determined. In quantum mechanics a similar explanation is provided by the Copenhagen interpretation and, more explicitly, by the model in Ref. \cite{topdown2}.

In fact, the latter mechanisms are strongly related to what has been discussed at length in the context of quantum theory, namely the long-standing ``quantum measurement problem''. This comprises the problem of ``explaining why a certain outcome --as opposed to its alternatives-- occurs in a particular run of an experiment'' \cite{brukner}. In fact, some of the most commonly accepted interpretations of quantum mechanics (e.g. the Copenhagen interpretation) uphold the view that it is the act of measurement to impose to microscopic (quantum) objects to actualize one determined value, out of the possible ones.

Note that, despite it has been already remarked in the literature that every indeterministic theory has to deal with a ``measurement problem'' (see e.g. \cite{brukner}), it seems that there has hardly been any consideration of this issue in the context of other indeterministic theories than quantum mechanics. In the next subsection we will discuss what we call, by analogy, the ``classical measurement problem'. We will then draw a connection with top-down causation, and show how this could help to shed light on this problem.

 \subsection{The ``classical measurement problem''} 
It is a very corroborated experimental fact that if a quantity is measured twice with the same instrument, we expect a certain amount of digits to remain unchanged, and it is essential to scientific investigation that such a knowledge is intersubjectively available (up to the digit corresponding to the measurement accuracy). Moreover, if a more accurate measurement instrument is utilized, we expect not only the previous digits to remain unchanged, but also to determine some new digits that then become intersubjectively available. How to reconcile this stability of the measured digits with a fundamental uncertainty in the determination of a physical quantity? How does potentiality  become actuality?
% (in the FIQ-based model here proposed, in the form of propensities)

In the proposed FIQ-based indeterministic interpretation of classical physics, too, one has to carefully define how the digits of physical quantities realize themselves from the propensity of taking that (or another) possible value. Consider for example the chaotic systems analyzed in \cite{gisin1} (a simplified version of the \emph{baker's map}). One can then think of a ``faster'' dynamics that, at every time step, shifts the bits by not only one digit toward the more significant position, but, say, it shifts them by 1000 digits (or any other arbitrarily large finite number). This clearly entails that the rate of change of propensities depends on the dynamical system under consideration, and cannot be thought of as a universal constant of ``spontaneous'' actualization.  A possible solution is to introduce a model of measurement that makes the digits becoming actual (and therefore stable) up to the corresponding precision. This clearly resembles the solution to the quantum measurement problem provided by the \emph{objective collapse models}, such as the CSL \cite{gisincollapse, cls} or the GRW \cite{grw, belljumps}. The latter model, indeed, posits a modification of the standard Schr{\"o}dinger's equation which accounts for a spontaneous random ``collapse'' of the wave function, occurring with a certain natural rate. Under certain assumptions, this model leads to a mechanism that changes the rate of spontaneous collapse, which increases linearly with the number of components of a system (thus during a measurement the wave function of a microscopic system in contact with a macroscopic apparatus collapses extremely fast). An analogous solution can be in principle proposed for the rate of actualization of the propensities that define the FIQs. However, this seems to mean that the dynamical equations need to be modified (in the same fashion as the GWR model modifies the Schr{\"o}dinger's equation), thus leading to a different formalism and not only an interpretation.

Coming back to top-down causation, this could explain why every time one performs a measurement the determined digits remain stable. In fact, the act of a measurement can be regarded as the direct action performed at the higher level which imposes to the lower level to get  determinate. This is very similar to what is taken to be the solution to the quantum measurement problem within the Copenhagen interpretation, wherein the higher level is the macroscopic measurement apparatus, whereas the lower level is the measured microscopic system. However, this kind of solutions lacks a clear definition of what is to be considered a measurement and how to identify higher and lower levels of description.

%Of course, one can conceive a solution to the quantum measurement problem that features a mixture of the previous two. In fact, it could be thought that the digits of physical quantities get realized spontaneously as time passes, with a certain natural rate, but when a higher order requires it, more digits are forced to get actualized (top-down causation). 

As a matter of fact, the ``classical measurement problem'' here introduced remains so far unresolved, as well as the quantum measurement problem and, more in general, the problem of the actualization of physical variables in any indeterministic theory. Yet, it is desirable that the topic of the measurement problem should find room in the debate on foundations of physics, in more general discussions than those centered on quantum mechanics only.

\section{Conclusions}

We have discussed arguments --primarily based on the modern application of information theory to the foundations of physics-- against the standard view that classical physics is necessarily deterministic. We have also discussed concrete perspectives to reinterpret classical physics in an indeterministic fashion. We have then compared our indeterministic proposals with some interpretations of quantum physics. However, it seems clear that the empirical results of both classical and quantum mechanics can fit in either a deterministic or indeterministic framework. Furthermore, there are compelling arguments (see e.g., \cite{gisin1, suppes, wendl}) to support the view that the same conclusion can be reached for any given physical theory --a trivial way to make an indeterministic theory fully determined is to ``complete" the theory with all the results of every possible experiments that can be performed.

In conclusion, although the problem of determinism versus indeterminism is in our opinion central to science, the hope to resolve this problem within science itself has faded, and this is ultimately to be decided on the basis of metaphysical arguments.

\subsection{Acknowledgments}
We would like to thank the participants to the Workshop ``Experiencing  Reality Directly: Philosophy and Science'', held in Jerusalem on May 20-22, 2019, for many discussions that provided interesting inputs. We also thank George Ellis, Veronica Baumann, Arne Hansen and Stefan Wolf for useful comments. FDS acknowledges the financial support through a DOC Fellowship of the Austrian Academy of Sciences.

\begin{small}

\end{small}

\end{document}